\documentstyle[12pt,epsf,aaspp,flushrt]{article} 
\begin{document}

   \title{Extragalactic Gamma-Ray Absorption and the Intrinsic Spectrum of 
Mkn 501 during the 1997 Flare.}

   \author{O.C. de Jager}
          
   \affil{Unit for Space Physics, Potchefstroom University 
for CHE, Potchefstroom 2520, South Africa \\}
   \author{F.W. Stecker}
\affil{Laboratory for High Energy Astrophysics, NASA Goddard Space
                Flight Center, Greenbelt, MD 20771, USA \\}



   \begin{abstract}
Using the recent models of Malkan \& Stecker (2001) for the infrared
background radiation and extrapolating them into the optical and UV range using
recent galaxy count data, we rederive the optical depth of the Universe to
high energy $\gamma$-rays as a function of energy and redshift for energies
betweeen 50 GeV and 100 TeV and redshifts between 0.03 and 0.3. 
We then use these
results to derive the intrinsic $\gamma$-ray spectrum of Mkn 501 during its 
1997 high state. We find that the time averaged
spectral energy distribution of Mkn 501 while 
flaring had a broad, flat peak in the $\sim$5--10 TeV range which corresponds
to the broad, flat time averaged
X-ray peak in the $\sim$ 50--100 keV range observed during the flare.
The spectral index of our derived intrinsic differential photon spectrum for
Mkn 501 at energies below $\sim$2 TeV was found to be 
$\sim$1.6--1.7. This corresponds to a time averaged spectral
index of 1.76 found in soft X-rays 
at energies below the X-ray (synchrotron) peak. These results appear to 
favor a synchrotron-self Compton (SSC) origin for the TeV emission together 
with jet parameters which are consistent with time variability constraints 
within the context of a simple SSC model.

   \end{abstract}

%

\section{Introduction}

The spectra of high energy $\gamma$-rays from extragalactic sources
are predicted to be modified by strongly redshift dependent absorption effects 
caused by interactions of these $\gamma$-rays with photons of the 
intergalactic IR-UV 
background radiation (Stecker, De Jager 
\& Salamon 1992). Several attempts have been
made to infer the IR SED (spectral energy distribution) either from model
calculations or observations. 
(See Hauser \& Dwek (2001) for the latest review.) Such information 
can be used to calculate the optical depth for TeV range photons as a function
of energy and redshift  
so that the intrinsic spectrum of an observed source can be derived.

Of particular interest is the spectrum of Mkn 501 which was observed while
strongly flaring in 1997. The spectrum observed at that time by the 
{\it HEGRA} air 
Cherenkov telescope system (Aharonian, {\it et al.} 1999, 2001a) 
extended to energies greater than 20 TeV, the highest energies yet observed
from an extragalactic source.

Application of the Stecker \& de Jager (1998)
calculations for the optical depth to the observed spectrum as first reported
reported by Aharonian, {\it et al.} (1999) resulted
in an intrinsic spectrum which was consistent with a differential power-law 
form for the photon spectrum
$\propto E^{-2}$ between 0.5 and 20 TeV, corresponding to a flat spectral
energy distribution (SED), {\it i.e.}, $E^2[dN/dE] \simeq$ constant 
(Konopelko, Kirk, Stecker \& Mastichiadas 1999). 
In this paper, we will rederive the intrinsic spectrum of Mkn 501 using the
method employed by Konopelko, {\it et al.} (1999) by using the recalibrated
observed spectrum as given by Aharonian {\it et al.} (2001a)
and using a consistent hybrid model for the extragalactic
background radiation. We correct for
absorption by first recalculating the opacity of intergalactic
space to $\gamma$-rays, $\tau(E,z)$, using the 
new calculations of the IR background spectrum of Malkan \& Stecker (2001)
extended into the optical domain as described later in this paper. 
The consistency between the {\it Whipple} telescope and 
{\it HEGRA} spectra of Mkn 501 allows us to derive
the intrinsic spectrum of Mkn 501 over two decades of energy.

The intrinsic TeV $\gamma$-ray SED which we derive here,  
$\nu F(\nu) = E^2[dN/dE]$, is again quite flat. However, it is slightly
convex, with a broad peak in the multi-TeV range.
The spectrum can be interpreted
as a Compton peak corresponding to a synchrotron peak which has been observed
in the X-ray range. This allows us to determine 
jet parameters which may explain the variability and spectrum of
Mkn 501. We discuss the implications of this spectrum 
in the context of the synchrotoron self-Compton source hypothesis 
(Stecker, de Jager \& Salamon 1996) and 
for constraints on the breaking of Lorentz invariance.

\section{A Hybrid SED for the IR-optical Intergalactic Radiation Field}

In order to recalculate the $\gamma$-ray opacity of intergalactic space
from photon-photon pair production interactions, we adopt a hybrid
model which uses the near-IR to far-IR background SED calculated
by Malkan \& Stecker (2001) combined with a reasonable, smooth extention 
into the optical-UV wavelength range derived from recent HDF (Hubble Deep 
Field) galaxy counts by Madau \& Pozzetti (2001). We will refer to the entire 
FIR-UV background as the EBL (extragalactic background light).

Malkan \& Stecker (2001) have used their empirically based model (Malkan \&
Stecker 1998) to predict infrared galaxy luminosity
functions and deep infrared galaxy counts at various wavelengths.
They have also examined their predictions for the IR SED for comparison with 
the subsequent determinations from the {\it COBE-DIRBE} data analysis.
Using the assumption of pure luminosity evolution proportional to $(1+z)^Q$ 
out to a redshift of $z_{flat}$ and constant (no evolution) for 
$z_{flat} < z < z_{max} = 4$,
they find that a comparison of their predictions with current {\it ISO} 
galaxy counts at 15 and 175 $\mu$m favor their ``baseline model" with $Q = 
3.1$ and $z_{flat} = 2$ (the lower curve in Figure 1). The mid-IR $\gamma$-ray 
upper limits of $\sim$ 4--5 nW m$^{-2}$sr$^{-1}$ 
(Stecker \& De Jager 1997; Stanev \& Franceschini 1998; 
Renault, {\it et al.} 2001) also favor $Q \sim 3$. 

On the other hand, the {\it COBE-DIRBE} 
far infrared background flux determinations of Hauser {\it et al.} (1998)
seem to favor a stronger 
evolution with  $Q > 4$ up to $z_{flat} = 1$.\footnote{There was a tentative
derivation of the background at 60 $\mu$m and 100 $\mu$m by Finkbeiner, 
Davis \& Schlegel (2001). However, this result now appears to have suffered 
from contamination by local solar system dust emission (Finkbeiner 2001).}
The upper curve in Figure 1, which shows the ``fast evolution'' model of
Malkan \& Stecker (2001), assumes $Q = 4.1$ and $z_{flat} = 1.3$.
The lower curve in Figure 1 shows the baseline model.

If one rederives the 140 $\mu$m and 240 $\mu$m background
fluxes from {\it COBE-DIRBE} using the {\it COBE-FIRAS} calibration, which
suffered from smaller systematic errors than the {\it COBE-DIRBE} calibration
(Fixsen {\it et al.} 1997), the fluxes are lowered to the point where they
are {\it consistent with both} the baseline and fast evolution 
SEDs of Malkan \& Stecker (2001) (see Figure 1). The newly derived flux
at 170 $\mu$m obtained from {\it ISOPHOT} maps by Kiss {\it et al.} (2001),
also shown in Figure 1, is also consistent with the both Malkan-Stecker
SEDs. In this regard, one should also note that the results reported by 
Hauser {\it et al.} (1998) were at the 4$\sigma$ level. 
(See Malkan \& Stecker (2001) for further discussion.)

Our hybrid approach extends the IR SEDs of Malkan \& Stecker (2001) to
shorter wavelengths by exploiting the fact that the Hubble Deep Field
(HDF) was able to resolve the lowest intensity
galaxies. In other words, a graph of the integral number of galaxies above a
given intensity as a  function of that intensity shows 
convergence to a constant value, which implies that the total amount
of extragalactic background light (EBL) 
has been captured in the wavebands of interest.
Thus, unless there are significant non-stellar, intergalactic contributions
to the optical background or very low surface brightness galaxies missed by
the {\it HST} (Hubble Space Telescope), the flux derived from the HDF should
represent the true EBL. 

Galaxy counts provide a strict lower limit on the
intensity of the EBL, since faint or unresolved galaxies not included in the 
counts are additional contributors to the EBL. The relative importance 
of the unresolved sources compared to that of the resolved
sources can be estimated from fluctuation analyses of the EBL. 
A lack of fluctuations will suggest a certain degree of
completeness in the source counts. Based on such fluctuation analyses, the
intensities reported by Madau \& Pozzetti (2001) can be considered 
as actual detections of the EBL at the wavelengths studied. Their results
are shown as such in
Figure 1. These authors also give a lower limit of 15 nW m$^{-2}$sr$^{-1}$ 
for the integrated light between 0.2 and 2.2 $\mu$m. 
\footnote {There is evidence for an upturn in the SED at the shortest 
observed UV wavelengts near $2\times 10^{15}$ Hz as derived from the {\it HST}
imaging spectrograph (Gardner, Brown and Ferguson 2000), but this feature does
not significantly affect our absorption calculations; the UV background only 
affects $\sim 50$ to 100 GeV $\gamma$-rays from sources at high redshifts 
(Salamon \& Stecker 1998).}    

The HDF K--band (2.2 $\mu$m)
flux is further complemented by ground based observations.
Gardner (1996) gives a lower limit of 
7.4 nW m$^{-2}$ sr$^{-1}$ on the EBL intensity, but deeper
K-counts (Bershady, Lowenthal \& Koo 1998)
suggest an increase in the K $>$ 23
counts over previous surveys (see {\it e.g.} Gardner 1996). 
The slope in the K-band counts, $dN/dm$ is
about 0.36. Adopting a slope of 0.4 down to K = 30 magnitudes, 
E. Dwek (1999, personal communication) derived a conservative upper limit of 
13 nW m$^{-2}$ sr$^{-1}$ for the EBL intensity in the K-band. 
These two limits are shown as a thick vertical bar in Figure 1 and provide
an important constraint when attempting to connect the far-IR to UV
parts of the EBL in any hybrid model. From Figure 1 we also see that the 
ground based limit on the K-band flux is consistent with the {\it HST} galaxy 
counts in the HDF. Dwek and Arendt (1998) have used {\it COBE-DIRBE} 
data to derive a tentative
estimate of the EBL flux at a wavelength of 3.5 $\mu$m of 9.9 $\pm$ 5.8 
(2$\sigma$) nW m$^{-2}$sr$^{-1}$, shown by the solid diamond point in Figure 
1, which is also seen to be in good agreement with our models.

It can be seen from Figure 1 that the Malkan \& Stecker (2001) models 
connect smoothly with the near-IR to UV lower limits for the EBL based
on HDF galaxy counts (Madau \& Pozzetti 2001). This fact allows us to 
construct a reasonable {\it hybrid model}, smoothly extending the infrared 
model SEDs of Malkan \& Stecker (2001) to UV wavelengths.

The thick curves in Figure 1 show the SEDs predicted by the baseline 
(lower curve) and fast evolution models (upper curve) of Malkan \& Stecker 
(2001). The baseline SED of Malkan \& Stecker (2001) connects
smoothly with the mean stellar component (open circles) derived
from galaxy counts ({\it HST} and ground--based), which is shown as the thin
dashed line (lower curve in Figure 1). The integral of this spectrum (extending
through the mean UV to K--band fluxes) between 0.2 $\mu$m and 2.2 $\mu$m
gives a value of 16 nW m$^{-2}$sr$^{-1}$, which is just
above the galaxy count lower limit of 15 nW m$^{-2}$sr$^{-1}$ derived
by Madau \& Pozzetti (2001). The fast evolution model of Malkan \& Stecker
(2001) also runs smoothly through the $2\sigma$ galaxy count limits
and gives us a reasonable upper limit for the EBL in the UV to far-IR.

Figure 1 summarizes the observationally derived values for the optical-UV, 
near-IR and far-IR fluxes which now exist. Unfortunately, foreground 
emission prevents the direct detection of the EBL in the mid-IR wavelength
range (See discussion in Hauser \&
Dwek 2001.) However, we note that other theoretical models such as those of 
Tan, Silk \& Balland (1999), Rowan-Robinson (2000) and Xu (2000) predict 
fairly flat SEDs in the mid-infrared range with average flux levels in the 
3 to 4 nW m$^{-2}$sr $^{-1}$ as do the Malkan \& Stecker (2001) models used 
here. These flux levels are also consistent with the indirect mid-IR 
constraints indicated by the box in Figure 1 (These constraints are summarized
by Stecker (2001).
Fine detailed differences between these theoretical SEDs are not 
important in determining the optical depth of intergalactic space to high
energy $\gamma$-rays, $\tau(E,z)$, (see next section) because its 
determination involves integrating over a significant range of EBL 
wavelengths (Stecker, {\it et al.} 1992). As an example of this, we note that
Vassiliev (2000) has shown that for a flat SED, 90\% of the contribution to
the $\gamma$-ray optical depth is contributed by EBL photons of wavelengths 
between .28 and 2.7 times the optimum wavelength corresponding to the peak 
in the pair production cross section. This is roughly an order of magnitude 
in integrated EBL wavelengths.

We therefore assert that the two Malkan \& Stecker curves,
extended into the optical-UV range by our hybrid model, give a reasonable 
representation of the EBL in the UV to far-IR. Other EBL models in the 
literature whose flux levels roughly fit the present data and have the same 
spectral characteristics, {\it i.e.}, a stellar optical peak, a far-IR dust 
emission peak, and a mid-IR valley which allows for some warm dust emission
(see review of Hauser \& Dwek 2001), 
should give similar results on the optical depth of the near Universe 
$\tau(E,z)$ to high energy $\gamma$-rays (see next section). 

\section{The Optical Depth for Very High Energy $\gamma$-rays from 
sources at $z<0.3$.}

We used the prescription of Stecker \& de Jager (1998), together with
the hybrid model discussed in the previous section, to recalculate 
the optical depth of intergalactic space $\tau(E,z)$ for redshifts between 
0.03 (near the $z$ values of
Mkn 501 and 421) and 0.3, assuming that the EBL is in place by a time
corresponding to $z = 0.3$.
Above $z\sim 0.3$, evolutionary effects
have to be taken into account (Salamon \& Stecker 1998).
Figure 2 shows $\tau(E,z)$ calculated using the 
baseline and fast evolution models for $\gamma$-ray energies
down to $\sim 50$ GeV, which is the approximate threshold energy for 
meaningful image analyses in next generation ground based 
$\gamma$-ray telescopes such as {\it MAGIC, H.E.S.S.} and {\it VERITAS}. 
Where our results
overlap, they are in good agreement with the metallicity corrected results
of Salamon \& Stecker (1998). We note that the Salamon \& Stecker (1998)
results extend to both lower energies and higher redshifts.

As a quantitative example of the uncertainty in the predicted value of 
$\tau(E)$ at the Mkn 501 redshift of 0.031 produced by the uncertainty in the
EBL flux, we note that the 19\% difference between our two model 
curves at the mid-IR wavelength of 10 $\mu$m (see Figure 1) maps into a 
difference of 21\% in the optical depth at 10 TeV (see Figure 2). This, in 
turn yields a difference of 34\% in the predicted Mkn 501 intrinsic 
$\gamma$-ray flux at 10 TeV. The uncertainty in the energy above which the 
optical depth is greater than 1 is indicated in Figure 2. We find that 
$\tau(E) \ge 1$ for $E \ge 2.3$ TeV assuming the ``fast evolution'' model for 
the EBL and $\tau(E) \ge 1$ for $E \ge 3.2$ TeV for the baseline model EBL. 

As in Stecker \& de Jager (1998), we also
obtained parametric expressions for $\tau(E,z)$ 
(as shown in Figure 2),
but the accuracy was improved by increasing the order of the
polinomials. The expressions are of the form
(for $z<0.3$ and $0.1<E_{\rm TeV}<50$): 
\begin{equation}
log_{10}[\tau(E_{\rm TeV},z)]\simeq\sum_{i=0}^4a_i(z)
(\log_{10}E_{\rm TeV}+2)^i
\end{equation}
where the z-dependent coefficients are given by
\begin{equation}
a_i(z)=\sum_{j=0}^3a_{ij}(\log_{10}{z})^{j}.
\end{equation}

Tables 1 (baseline model) and 2 (fast evolution, numbers in
brackets)
give the numerical values for 
$a_{ij}$, with $j=0,1,2,3$, and $i=0,1,2,3,4$. 
Equation (2) approximates $\tau(E,z)$ correctly within 5\% 
for all bounded values of $z$ and $E$ considered.

\section{The Observed Spectrum of Mkn 501 during the 1997 Flare.}

Recently the {\it HEGRA} group (Aharonian {\it et al.} 2001a) 
rederived the energy spectrum of
Mkn 501 during the 1997 flare. The validity of the high energy
tail of the spectrum out to 20 TeV in their earlier analysis 
(Aharonian {\it et al.} 1999) was questioned because of 
spillover effects between energy channels. A reanalysis 
of the events above 3 TeV with better energy resolution gave
a flux at $\sim$ 21 TeV which was a factor of 3.6 
lower than that obtained in their previous analysis.
It should be stressed that their revised spectrum is 
consistent with the original spectrum below 20 TeV.
The final spectrum between 0.56 and 21 TeV is shown in Figure 3
and marked ``OBSERVED''.

Contemporaneous observations of Mkn 501 by the {\it Whipple}
group (Krennrich {\it et al.} 1999)
are consistent (within errors)
with the {\it HEGRA} spectrum in the overlapping energy
range between 0.5 and 10 TeV. This allows us to add the two {\it Whipple}
points below 0.5 TeV to the {\it HEGRA} spectrum, giving an observed 
spectrum between 0.26 and 22 TeV
for the Mkn 501 flare. The result is a single spectrum extending
over two decades of energy as shown in Figure 3.

\section{The Intrinsic VHE Spectrum of Mkn 501 during the 1997 Flare.} 

By using both the Whipple and {\it HEGRA}
spectra of Mkn 501 and correcting for absorption by multiplying
by $e^{\tau(E)}$ evaluated at $z = 0.034$
with the newly derived values for the opacity 
calculated as discussed in the previous section, we have derived  
the intrinsic spectrum of Mkn 501 over two decades of energy.
This is given by the data points and two curves marked INTRINSIC in
Figure 3. The upper of these curves corresponds to the fast evolution case;
the lower curve corresponds to the baseline model case.  

Fossati {\it et al.} (2000) 
suggested a parameterization to describe 
smoothly curving blazar spectra. This parameterization is of the form

\begin{equation}
\frac{dN}{dE}=
KE^{-\Gamma_1}\left(1+(\frac{E}{E_B})^f\right)^{(\Gamma_1-\Gamma_2)/f}
\end{equation}
A  spectrum of this form
changes gradually from a spectral index of $\Gamma_1$
to an index of $\Gamma_2$ when the energy $E$ increases through the
break energy $E_B$. The parameter $f$ describes the rapidity (``fastness'')
of the change in spectral index over energy. This model was used by
Fossati {\it et al.} (2000) to describe the curvature in the Mkn 421
spectrum. We have fitted our derived intrinsic Mkn 501 spectrum to 
the form given by equation (3).

We applied the formalism of Fossati {\it et al.} (2000) and found the 
best-fit parameters
$K$, $E_B$, $f$, $\Gamma_1$ and $\Gamma_2$ after correcting the
observed spectrum for intergalactic absorption. 
A total of 27 data points between 0.26 ad 21 TeV
were used for each of the two intrinsic
spectra in Figure 3, and we fitted to the Fossati {\it et al.}
parameterization of these spectra, keeping all parameters free.
Whereas the low energy spectral index
$\Gamma_1$ appears to be well constrained, we find that the
higher energy index $\Gamma_2$ is unconstrained.
Table 3 shows the parameter values for each fit, assuming
two arbitrary values for $\Gamma_2$ (2.5 and 3.0). The parameter $E_M$ is
derived from the fit, and corresponds to the $\gamma$-ray energy where
the SED peaks in $\nu F_\nu$. 
The fits are acceptable in terms of the $\chi^2$ per degree
of freedom. 

It is important to note that the intrinsic spectrum which we derive does not 
have an upturn near 20 TeV as obtained previously by some authors 
({\it e,g.} Protheroe \& Meyer 2000; Dwek 2001) working with the earlier 
analysis of Aharonian, {\it et al.} (1999) as well as the 60 $\mu$m flux
earlier reported by Finkbeiner {\it et al.} (2000 - see footnote 1).
The upturn obtained by Dwek (2001) can be effectively eliminated by 
using the revised {\it HEGRA} $\gamma$-ray
data with the observed flux reduced by a factor of 4 at $\sim$20 TeV; the
stronger upturn obtained by Protheroe \& Meyer (2001) is eliminated both by
using the revised HEGRA data and a more realistic IR-SED than the one these
authors used which had no mid-IR valley and an unreasonably high mid-IR flux.

Aharonian, Timokhin \& Plyasheshnikov (2001b) have also found an upturn
in their analysis, particularly if they use an SED constructed using the
flux claimed by Lagache {\it et al.} (2000) at 100 $\mu$m, the only point 
in Figure 1 which is inconsistent with our SEDs at the 2$\sigma$ level. 
Assuming that this upturn is correct, these authors suggested
avoiding a `` TeV-IR crisis'' by invoking a cold
jet near the central engine with a bulk Lorentz factor greater than $10^{7}$, 
with the upturn resulting from comptonization. 

In this context, we note that the {\it COBE-DIRBE} group has argued that
a real flux derived from the {\it COBE-DIRBE} data at 100 $\mu$m, as claimed 
by Lagache {\it et al.} (2000), is untenable because isotropy in the 
residuals (after foreground subtractions)
could not be proven. Dwek {\it et al.} (1998) have concluded that only a
conservative lower limit of 5 m$^{-2}$sr$^{-1}$ could be inferred at
100 $\mu$m. For Model II of Aharonian {\it et al.} (2001b) which 
neglects the 100 $\mu$m point, these authors derive a source spectrum which is 
{\it quite flat up to an energy of $\sim$ 20 TeV}, as we do in this work.
This obviates any need to invoke extreme bulk Lorentz factors in modelling the
Mkn 501 source spectrum.

Our intrinsic SED is slightly convex, rather than a single
power law, consistent with an SSC  
hypotheses for the origin of the TeV radiation.
We find that this SED peaks
at $E_{M} \sim 8-9$ TeV (independent of the unconstrained $\Gamma_2$)
in the case where the fast evolution EBL is assumed and that
$E_{M} \sim 5$ TeV if the baseline EBL is assumed (see Figure 3 and Table 3.)

\newpage

\section{A Simple Interpretation of the Intrinsic Mkn 501 Flare Spectrum 
within the Context of the SSC Paradigm}

If one observationally determines a flaring variability time $\Delta$t, the
optical depth for 
$\gamma$ -- $\gamma$ absorption from pair production is then given by
(Dermer \& Schlickheiser 1994; Aharonian, {\it et al.} 1999)
\begin{equation}
\tau_{\gamma\gamma} \simeq 0.055\phi_{10}\Delta t_{\rm day}^{-1}
\delta_{10}^{-6}H_{65}^{-2}E_{\rm TeV}
\end{equation}
using the normalizations $\delta_{10} \equiv \delta/10$ and
$\phi_{10} \equiv \phi_{\rm r}/10^{-10}$ erg cm$^{-2}$s$^{-1}$ with
$\phi(\epsilon) \equiv \epsilon^2(dN/d\epsilon)$  
being the observed energy flux of the low energy radiation.
The flux $\phi(\epsilon)$ is 
determined at $\epsilon \sim 100 \delta_{10}^2E_{\rm TeV}^{-1}$ eV which
is the energy range of the low energy photons responsible for the intrinsic 
optical depth of high energy $\gamma$-rays
in the jet. The value of the relevant low energy radiation flux 
$\phi_{10}$ is approximately 0.5 in the optical--UV range.
$H_{65}$ is the Hubble constant is in units of 65 km s$^{-1}$ Mpc$^{-1}$.

The time variation of the TeV $\gamma$-ray flux scales
as the square of the time variation of the X-ray flux. This scaling favours 
a synchrotron-self Compton origin
for the Mkn 501 source with the synchrotron peak in the X-ray range and
the Compton peak in the TeV $\gamma$-ray range (Krawczynski, {\it et al.}
2000). The X-ray SED of Mkn 501 appears to have
a maximum in the $\sim$ 50-100 keV range in the synchrotron component 
(Catanese {\it et al.} 1997; Pian {\it et al.} 1998; Petry, {\it et al.}
2000). This shape should be reflected in the high energy $\gamma$-ray 
domain, but with different
fastness ($f$) and slopes ($\Gamma_1$ and $\Gamma_2$) owing to the
energy dependence of the cross section for inverse Compton scattering
as the possibility of having more than one soft target photon
component for such scattering (Stecker, De Jager \& Salamon 1996). 

Since Mkn 501 is a giant elliptical 
galaxy with little dust, it is also reasonable to assume that the galaxy itself
does not produce enough infrared radiation to provide a significant opacity
to high energy $\gamma$-rays. Also, if our EBL is approximately
correct, it is reasonable to assume that the dominant absorption process is
intergalactic and that pair-production in the jet in negligible. This
hypothesis is supported by the fact that the high energy $\gamma$-ray SED
did not steepen during the flare.
This implies that the optical depth given by eq. (4)
is less than unity out to the highest observed energy $E \sim
20$ TeV.
The pair-production opacity constraints then limit the Doppler factor to
\begin{equation}
\delta_{10} \simeq
1.2\left(\frac{\phi_{-10}}{0.5}\right)^{1/6}
\left(\frac{5\;{\rm hours}}{\Delta t}\right)^{1/6}H_{65}^{-1/3}
\left(\frac{E_{\gamma}}{20\;{\rm TeV}}\right)^{1/6}
\end{equation}
which is consistent with the lower limit of $\delta_{10} = 1$ set
by Aharonian {\it et al.} (1999).

The energy of the synchrotron peak in the X-ray region is given by
\begin{equation}
E_{X,M}({\rm keV}) \simeq 67 B_{0.1}\delta_{10}[E_{e}({\rm TeV})]^2
\end{equation}
where $E_{e}$ is the energy of the electrons which are radiating into the
synchrotron peak and $B_{0.1}$ is in units of 0.1 G.
The energy of the Compton peak in the Klein-Nishina range is given by
\begin{equation}
E_{M} = \delta\kappa E_{e}
\end{equation}
where the inelasticity factor, $\kappa \sim 1$ in the Klein-Nishina limit.

From equations (4) and (5), we obtain the SSC expression for $E_{M}$,
\begin{equation}
E_{M}({\rm TeV}) \simeq 1.2\kappa\delta_{10}^{1/2}B_{0.1}^{-1/2}E_{X,M}({\rm keV})^{1/2}.
\end{equation}
Note that this expression also holds if SEC is important.
Krawczynski {\it et al.} (2000) have estimated the $B$ 
field from the observed lag in the cooling time between the 3 keV and 12 keV 
synchrotron radiation in the jet to be $\ge$ 0.025 G.

Taking $\delta_{10} \sim 1$, $B_{0.1} = 1$, $\kappa \sim 1$ and 50 keV $\le
E_{X,M} \le$ 100 keV (Pian, {\it et al.} 1998) then eq. (8)
gives a prediction of $E_{TeV,M} \sim$ 9 -- 12 TeV, in good 
agreement with the results of the $\chi^2$ fit to our intrinsic SED for the 
Mkn 501 flare of $\sim$ 5 -- 9 TeV. (Note that this value only depends on the 
square roots of the parameters involved.) 

It should be noted that Tavecchio,
{\it et al.} (1998) neglected intergalactic absorption and used the observed
spectrum of Mkn 501 with an assumed Compton peak at sub-TeV energies to 
determine constraints on SSC jet parameters. The fact that they failed to 
derive a self-consistent model shows the importance of taking intergalactic
absorption into account in theoretical work involving TeV blazars.

Our interpretation is based on {\it time averages} of flare 
spectra measured during the 1997 high state. However, our analysis is 
validated by the analyses of Petry {\it et al.} (2000).
These authors also compared the TeV data with the soft- and hard X-ray data
after averaging the X-ray data over several flares during the same
high state. Our comparison uses
both the soft X-ray spectral index and the hard X-ray peak taken from
the Petry {\it et al.} (2000) averages.

\section{Conclusions}

The IR SEDs which Malkan \& Stecker (2001) have derived and argued are
consistent with all of the reliable data on the IR EBL connect smoothly
with the EBL derived from converging galaxy counts 
in the near-IR to UV. Furthermore, at
the longest wavelengths, they also connect 
smoothly with the {\it COBE-FIRAS} results. 
This allows us to construct
hybrid EBL SEDs which were used
to recalculate the optical depth versus redshift for 
high energy $\gamma$-rays 
between 50 GeV and 100 TeV, the energy above which the Universe becomes 
highly opaque, even at low redshifts, owing to interactions with photons of 
the 2.7K cosmic background radiation (Stecker 1969).

We find that our derived intrinsic flaring SED for Mkn 501 exhibits
a broad maximum in the $\sim$ 5-10 TeV range, which is consistent with
a value of $\sim$ 10 TeV predicted from several consistent physical arguments
and independent observational constraints for the SSC model (see previous
section). 
The Mkn 501 SED derived in this paper does not show
the upturn above $\sim$ 20 TeV found in some other studies. We agrue here
that such a feature is an artifact produced partly
by spillover in the last data bin in the original {\it HEGRA} analysis and
partly by assuming too high an IR background SED.

Our differential photon spectral index, $\Gamma_{1}$, for the energy range 
$E_{\gamma}\ll E_{\rm M}$
is 1.6 to 1.7 (Table 3), which is close to the 
differential photon spectral index of 1.76 at soft X-ray energies
(Petry {\it et al.} 2000). The similarity of these spectral indices implies
that $\gamma$-rays below $\sim 1$ TeV are produced in the Thomson
regime by scattering off soft photons with energies in the 
optical-IR range. On the other hand, $\gamma$-rays near the $\sim$ 7 $\pm$ 
2 TeV Compton peak are the result of scattering in the Klein-Nishina range
(see previous section).

Our derived intrinsic Mkn 501 SED is quite flat in the multi-TeV range as 
shown in Figure 3. This is in marked contrast to the dramatic turnover in its 
observed SED. We argue that this is strong evidence that the
observed spectrum shows just the absorption effect predicted. Among other
consequences of this conclusion, the observed absorption feature allows one
to put strong constraints on the breaking of Lorentz invariance (Stecker \&
Glashow 2001). 

\begin{acknowledgements}
\noindent 
Acknowledgment: We would like to thank Felix Aharonian, Eli Dwek and Karl 
Mannheim for useful discussions.
\end{acknowledgements}

\newpage

\begin{table*}
\begin{center}
\caption{\bf Polynomial coefficients $a_{ij}$ for the 
baseline model.}
\begin{tabular}{crrrr}
\hline
$i$ & $a_{i0}$ & $a_{i1}$ & $a_{i2}$ & $a_{i3}$ \\ \hline
0&-5.1512&-4.4646&-3.5143&-0.80109\\
1& 9.2964&16.275&10.7540&2.51610\\
2&-4.8645&-12.86&-8.6982&-2.07510\\
3& 1.1164&3.9795&2.747&0.66596\\
4&-0.084039&-0.42147&-0.29644&-0.072889\\ \hline
\end{tabular}
\end{center}
\end{table*}

\begin{table*}
\begin{center}
\caption{\bf Polynomial coefficients $a_{ij}$ for the 
fast evolution model.}
\begin{tabular}{crrrr}
\hline
$i$ & $a_{i0}$ & $a_{i1}$ & $a_{i2}$ & $a_{i3}$ \\ \hline
0&(-5.4748)&(-4.7036)&(-3.5842)&(-0.79882)\\
1&(10.444)&(17.114)&(11.173)&(2.58430)\\
2&(-5.8013)&(-13.733)&(-9.2033)&(-2.17670)\\
3&(1.4145)&(4.3143)&(2.9535)&(0.71046)\\
4&(-0.11656)&(-0.46363)&(-0.32339)&(-0.078903)\\ \hline
\end{tabular}
\end{center}
\end{table*}

\begin{table*}
\begin{center}
\caption{Results of a multiparameter fit of a continuous
curvature model to the absorption corrected differential photon fluxes $dN/dE$
of Mkn 501 during its 1997 high state. Errors are typically
on the last digit.}
\begin{tabular}{lccccccc}
\hline
EBL  & $K$ ($\times 10^{-10}$) & $f$ & $E_B$ & $\Gamma_1$ & 
$\Gamma_2$ & $\chi^2$/dof & $E_M$ \\
 Model &(ergs cm$^{-2}$s$^{-1}$) & & (TeV) & & & & (TeV) \\
\hline
Fast Evolution    & 2.64 & 1.6 & 15.0 & 1.61 & 3.0 & 1.26 & 9.0 \\
Fast Evolution    & 2.62 & 2.1 &  8.8 & 1.62 & 2.5 & 1.24 & 7.9 \\
Baseline  & 2.30 & 2.3 &  9.0 & 1.70 & 3.0 & 1.32 & 4.7 \\
\hline
\end{tabular} 
\end{center} 
\end{table*}

\newpage
\section*{Figure Captions}

\noindent Figure 1. The SED of the EBL (see text for
references and descriptions). All error bars given at the $\pm 2\sigma$ 
level. 
The Malkan \& Stecker (2001) fast evolution model is shown by the upper, 
thick solid curve
between $\log_{10}\nu=11.8$ to 13.8) and their baseline model is shown by
the thick dashed line over the same frequency range.
Convergent Hubble Deep Field galaxy counts (Madau \& Pozzetti 2001): 
open circles; Ground--based galaxy counts limits at $2.2\mu$m 
(see text): thick vertical bar marked ``$2.2\mu$'';
{\it COBE-DIRBE} photometric sky residuals (Wright \& Reese 2000): small open squares;
TeV $\gamma$-ray-based upper limits: thick box marked ``TeV'' (see text for
references);
{\it ISOCAM} lower limit at $15\mu$m: upward arrow marked $15\mu$ (Elbaz
{\it et al.} 1999);
{\it COBE-DIRBE} far-IR sky residuals (Hauser {\it et al.} 1998): large open squares
at 140 $\mu$m and 240 $\mu$m;
{\it COBE-DIRBE} data recalibrated using the {\it COBE- FIRAS} calibration 
(see text): large solid
squares without error bars; {\it ISOCAM} 170 $\mu$m flux (Kiss {\it et al.}
2001): solid circle; 
{\it COBE-FIRAS} far-IR sky residuals (Fixsen, {\it et al.} 1998): 
thick dot-dash curve 
with $\sim 95\%$ confidence band (thin dot--dash band);
$100 \mu$m {\it COBE-DIRBE} point (Lagache {\it et al.} 2000): 
small solid square;
flux at 3.5 $\mu$m from {\it COBE-DIRBE} (Dwek \& Arendt 1998): solid diamond.

\vspace{1.0cm}

\noindent Figure 2. The optical depth for $\gamma$-rays above 50 GeV
given for redshifts between 0.03 and 0.3 (as labelled) calculated using the 
medium (dashed lines) and fast evolution (solid lines)
SED of Malkan \& Stecker (2001). 
 
\vspace{1.0cm}

\noindent Figure 3. The observed spectrum and derived intrinsic spectrum
of Mkn 501. The observed spectral data are as measured by
{\it HEGRA} (solid circles) and {\it Whipple} (solid squares). The upper points
are the absorption corrected data points (marked ``INTRINSIC'') using our
fast evolution hybrid EBL (upper data set and solid curve fit) and baseline
hybrid EBL (lower data set with dashed curve fit). The fit parameters are
given in Table 1.

\newpage

\begin{figure}
\epsfxsize=18cm
\epsfbox{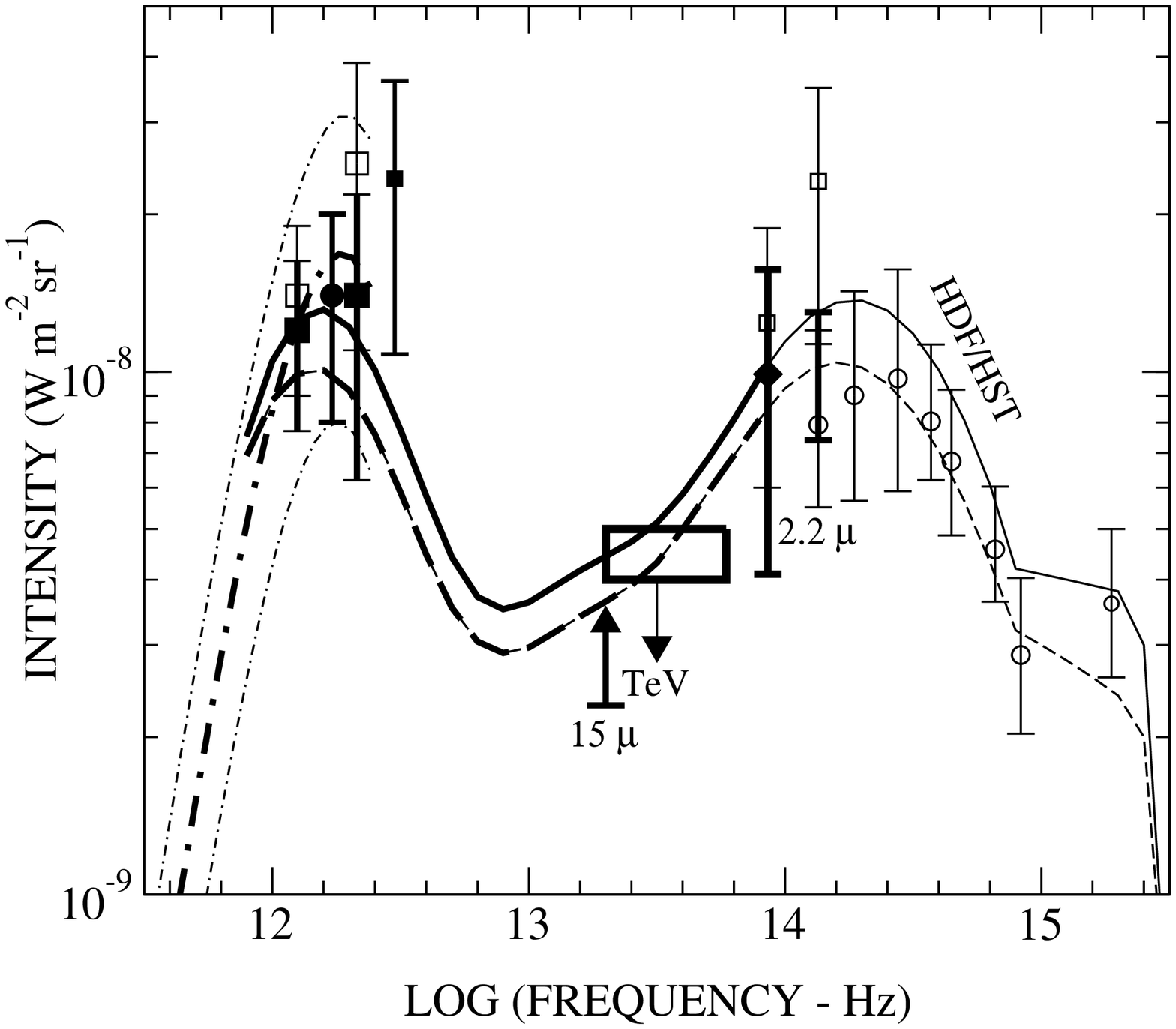}
\end{figure}

\newpage

\begin{figure}
\epsfxsize=18cm
\epsfbox{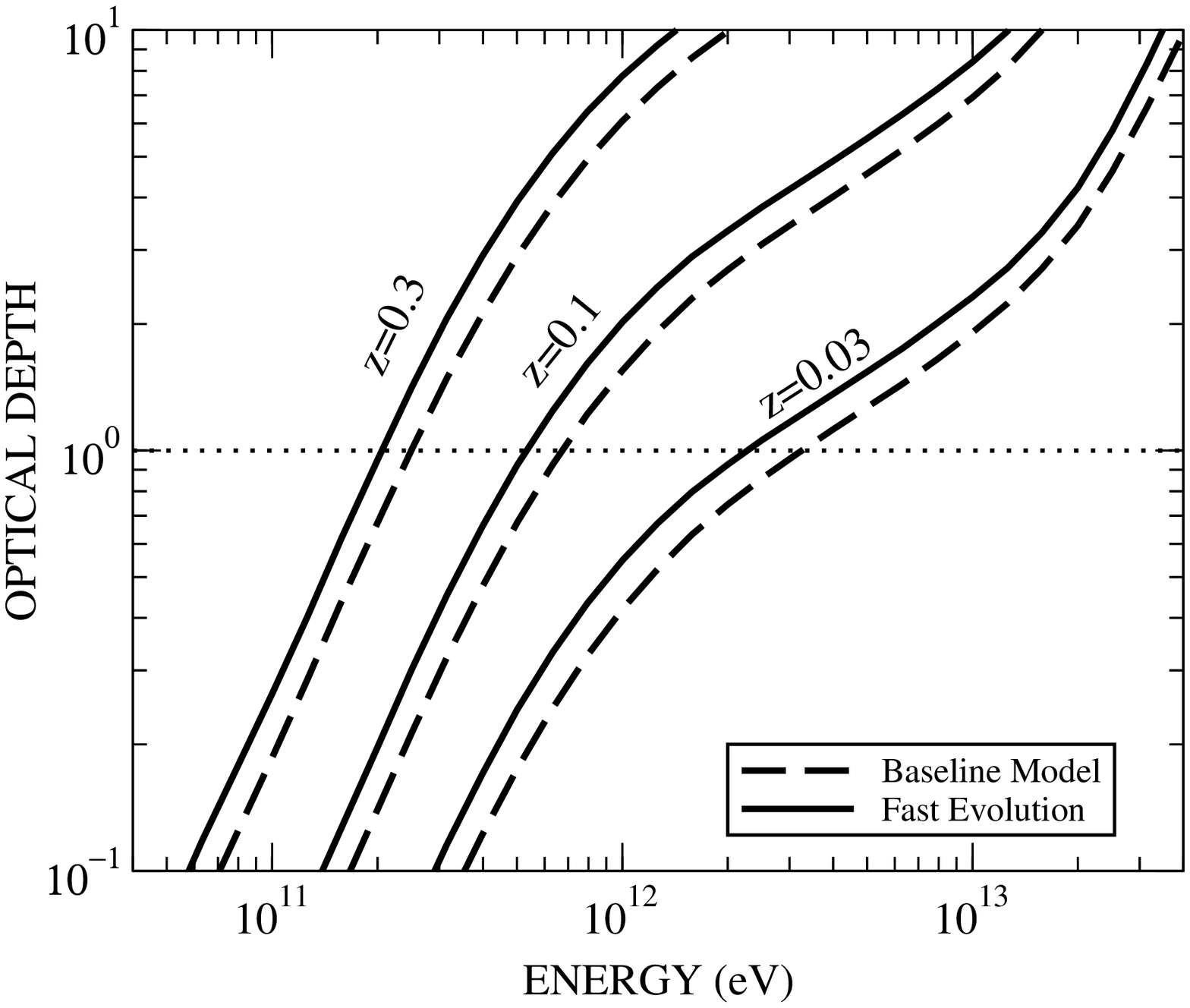}
\end{figure}

\newpage

\begin{figure}
\epsfxsize=18cm
\epsfbox{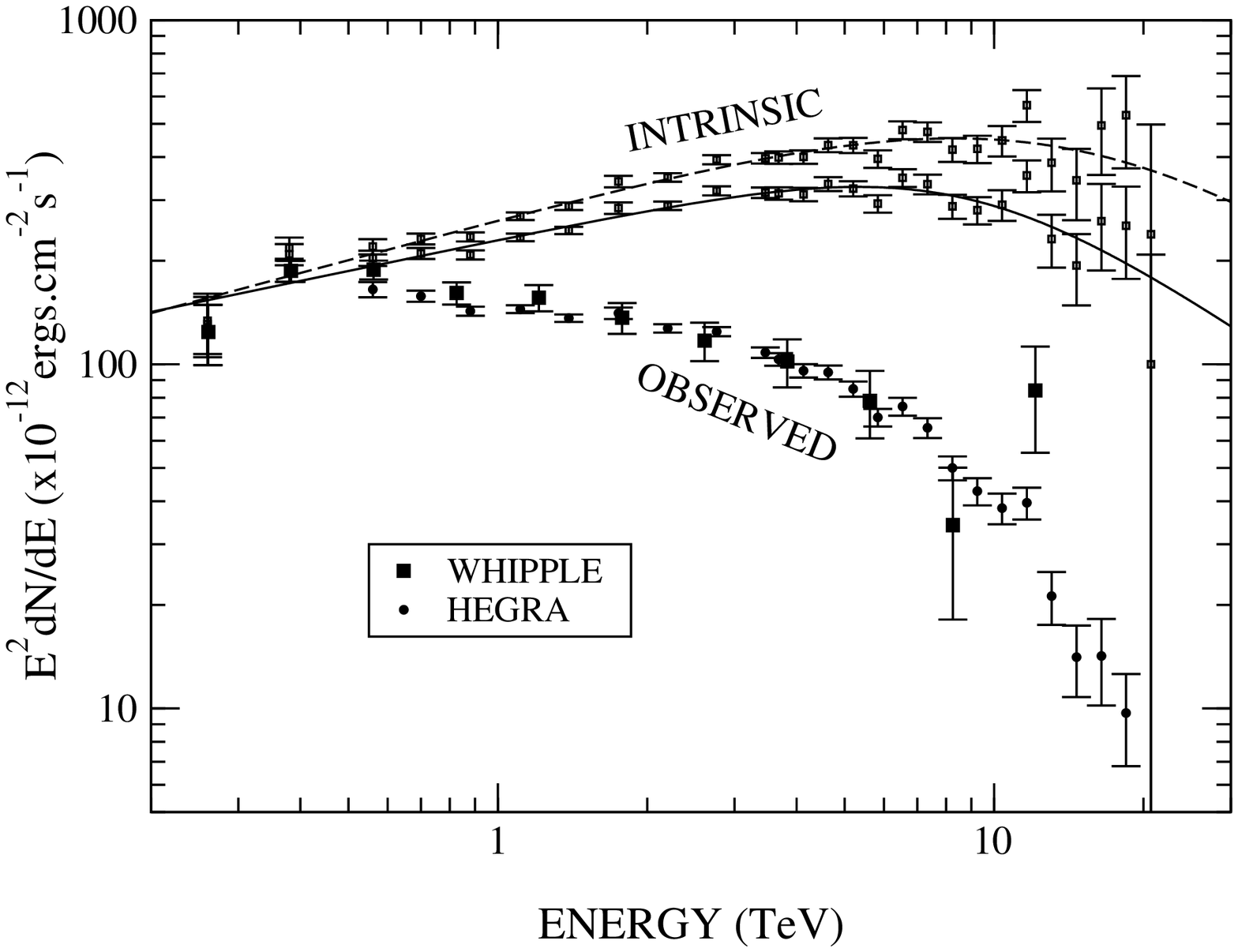}
\end{figure}

\end{document}